# Self-similar Fragmentation Regulated by Magnetic Fields in a Massive Star Forming Filament

Hua-bai Li[1], Ka Ho Yuen[1], Frank Otto[1], Po Kin Leung[1],
T. K. Sridharan[2], Qizhou Zhang[2], Hauyu Liu[3], Ya-Wen Tang[3], Keping Qiu[4]

**Most molecular clouds are filamentary or elongated[1,2,3]. Among those forming low-mass stars (< 8 $M_\odot$), their long axes tend to be either parallel or perpendicular to the large-scale (*10-100 pc*) magnetic field (B-field) in the surrounding inter cloud medium[3]. This arises because, along the dynamically dominant B-fields, the competition between self-gravity and turbulent pressure will shape the cloud to be elongated either perpendicular[4] or parallel[5] to the fields. Recent study also suggested that, on the scales of *0.1-0.01 pc*, fields are dynamically important within cloud cores forming massive stars (> 8 $M_\odot$)[6]. But whether the core field morphologies are inherited from the inter cloud medium or governed by cloud turbulence is under vigorous debate, so is the role played by B-fields in cloud fragmentation at *10 - 0.1 pc* scales[7,8,9]. Here we report B-field maps covering *100-0.01 pc* scales inferred from polarimetric observations of a massive-star forming region, NGC 6334. First, the main filament also lies perpendicular to the ambient field. NGC 6334 hosts young star-forming sites[10,11,12] where fields are not severely affected by stellar feedback, and their directions do not change significantly over the entire scale range. This means that the fields are dynamically important. At various scales, we find that the hourglass-shaped field lines are pinched where the gas column density peaks and the field strength is proportional to the 0.4-power of the density. We conclude that B-fields play a crucial role in the fragmentation of NGC 6334.**

At a distance of ~1.7 kiloparsecs, NGC 6334 is one of the nearest massive-star forming regions. Other massive-star forming sites are usually too far away to effectively use star-light polarization (due to extinction from dust grains aligned by B-fields) to probe B-field orientations (the two directions are parallel) in the surrounding inter cloud medium. B-field directions can be derived from the polarization of local background stars subtracting the polarization of local foreground stars[13]. To form a typical giant molecular cloud, gas needs to be accumulated from an inter cloud medium of a few hundred parsecs[14]. Using the optical polarimetry archive from Heiles[15], the ambient B-field direction of NGC 6334 is seen to be perpendicular to its elongation (Fig. 1).
If the B-field is dynamically important compared to turbulence during the gas accumulation process, the ambient B-field direction should be preserved inside the cloud[8]. Massive-star form-ing regions have stronger thermal dust emissions than their low-mass counterparts, which allows probing of the B-fields within dense clouds using polarization of submillimeter thermal dust emissions (the two directions are perpendicular). This has been performed extensively[13,16,17], but never with such a wide range in scales as presented here. We use data acquired by the polarimeters SPARO (10-parsec scale)[13], Hertz (parsec scale)[18], and the Submillimeter Array (SMA, 0.1-parsec scale)[6]. Following reference 19, interpolation of independent 3-sigma polarimetry detections are used to plot the B-field lines (Figs. 1-2).

[1] Department of Physics, The Chinese University of Hong Kong, Shatin, NT, Hong Kong SAR

[2] Harvard-Smithsonian Center for Astrophysics, 60 Garden Street, Cambridge, MA 02138, United States

[3] Academia Sinica Institute of Astronomy and Astrophysics, 11F of Astronomy-Mathematics Building, AS/NTU No.1, Sec. 4, Roosevelt Rd, Taipei 10617, Taiwan, R.O.C.

[4] School of Astronomy and Space Science, Nanjing University, 22 Hankou Road, Nanjing, Jiangsu, 210093 P.R.China

It is obvious that the field lines at 10-parsec scale are *"pinched" near the ends of the dust filament, where the massive-star-forming clumps, I/I(N) and IV, are also located* (Fig. 1). Hertz resolved the density peaks, showing that *I and I(N) are again situated near field-line pinches at the parsec scale*.
SMA further zoomed in onto the density peaks of I, I(N), and IV. I(N) is the youngest among the three cores, with weak outflows[6,10] and a low temperature (~30 K)[11]; the field lines are again *symmetrically pinched*. The more developed core I is hotter (~100K)[12], and has high-velocity outflows in the NE-SW direction[10], which might have altered the field direction from the larger scale. The curved filament of IV is part of a compression shell due to the $H_{II}$ bubble[20], which compressed the B-fields at the same time; hence the field and filament are largely aligned.

The average orientations of the filamentary cloud, elongated clumps/cores and the B-fields (defined by "equal weight Stokes mean"[13]) are summarized in Fig. 3. The orientation of a cloud is defined by the long-axis direction of the autocorrelation function of the intensity map[3]. There are several intriguing facts revealed by Fig. 3. First, assuming turbulence is the only force that disturbs B-field orientations and has the same energy density as the B-fields, the dispersion of B-field directions should be 30° based on the Chandrasekhar-Fermi relation[3,21,22]. Apart from region IV, all the field orientations in Fig. 3 are within this 30° range. In reality, field dispersions are not only due to turbulence, but also gravity[16,17], stellar feedback (e.g., region IV) and projection, so the turbulent energy of NGC 6334 should be sub-Alfvenic (*Method 1*).
Secondly, at all the scales, we observed hourglass-shaped or ordered B-fields to be close to perpendicular to cloud elongations unless severely affected by stellar feedback (core IV). This is a signature of the Lorentz force supporting the cloud against gravitational contraction in the direction perpendicular to the field lines[16,17]. This anisotropic contraction will result in flattened structures, which will appear elongated and tend to be perpendicular to the B-field projection[3]. Third, a thin sheet should fragment at the rim instead of at the center of the mass due to the difference in gravitational contraction velocity over the sheet[23]. The sky projection should appear as off-centered density peaks near the field-line pinches, which is also observed at multiple scales.

How B-field strengths vary with gas density also tells the role of B-fields in the contraction of cloud fragments. If B-fields are dynamically unimportant, the contraction should be *isotropic*, which results in $B \propto n^{2/3}$, where *n* is density. This is because contraction along the B-field direction can only enhance *n* but not *B*. The exponent should be less than 2/3 if B-fields are strong enough to channel the contraction to some extent[24,25]. Since we cover multiple scales and thus multiple densities, we can study this dependence and its implications.

We can estimate the B-field strength based on the balance between the forces from gravity ($F_G$), magnetic pressure ($F_P$) and magnetic tension ($F_T$) (Fig. 4; *Method 2*). The $F_G$ between the density peaks is:

$$F_G = 2.8 \times 10^{28} (\frac{M_1}{100 M_\odot})(\frac{M_2}{100 M_\odot})(\frac{0.1 pc}{D})^2 \; dyn,$$

where $M_1$ and $M_2$ are the masses of the two dense clumps; $D$ is the distance between the peaks. Presenting B-field orientations as field lines[19] allows us to estimate field line curvatures and thus $F_T$:

$$F_T = \frac{V}{4\pi} B \cdot \nabla B \sim \frac{V}{4\pi} \frac{B^2}{R} = 1.5 \times 10^{30} (\frac{B}{1mG})^2 (\frac{0.5 pc}{R})(\frac{V}{1pc^3}) \; dyn,$$

where $R$ is the radius of the field-line curvature and $V$ is the clump volume. Field lines near the density peaks and with more prominent curvatures are selected (noted in Fig. 1 & 2) to estimate $R$. This will give a lower-limit of $R$ (and thus $B$).

If there exists a gradient of the B-field strength, $F_P$ should also be considered:

$$F_P = -\nabla(\frac{B^2}{8\pi})V \sim \frac{V}{8\pi} \frac{B^2 - B_0^2}{r/2} = 1.5 \times 10^{30} \left[ (\frac{B}{1mG})^2 - (\frac{B_0}{1mG})^2 \right] (\frac{0.5pc}{r})(\frac{V}{1pc^3}) \; dyn,$$

where $B_0$ is the field strength outside the hourglass and $r$ is twice the "waist" (approximated by $D$). At the 10-parsec scale, the Galactic field strength, 10 µG[25], is used for $B_0$. Estimates of $B$ on the 10- and 1-parsec scales are used as $B_0$ for the 1- and 0.1-parsec scales respectively. $M$, $D$ and $V$ can be derived from the literature[26,27] (Extended Data Table I) and $R$ is measured from our maps (Figs. 1, 2 and 4).

Setting $F_G = F_P + F_T$ (*Method 2*) results in $B$ of approximately 0.2, 1.2 and 12 mG on the 10, 1 and 0.1 parsec scales respectively (Extended Data Table I). Note that the line-of-sight (LOS) and the Sagittarius spiral arm (and thus the Galactic B-field) are almost perpendicular at the position of NGC 6334. Hence the projection effect on the field curvatures and thus strengths should not be severe. The LOS component of $B$ at parsec scale measured by Zeeman effect is roughly 0.2 mG[28], which implies that the angle between the B-field and LOS is 80°.

Finally, approximating $n$ by $(M_1 + M_2)/D^3$ yields $B \propto n^{0.41 \pm 0.04}$ (Extended Data Fig. 1), with an exponent significantly lower than 2/3. This is the first $B$-$n$ relation derived from one single cloud covering 10-0.1 pc. Previously, the exponents are mainly based on Zeeman measurements where different $n$'s are obtained from different types of clouds that do not have any connection[24,25]. We can further show that the mass-to-magnetic flux ratios of the cloud/clump/cores are on average 1.6±0.5 relative to critical (*Method 3*). This agrees with the value required to form massive stars in recent numerical studies[9].

The magnetic topology problem[29], i.e., how the field topology evolves as molecular clouds form out of the ISM and as cores contract to form stars, has puzzled astronomers for decades, largely due to the difficulties on observations. After a decade of data collection, we finally shed some light on this problem. The Atacama Large Millimeter/submillimeter Array will have adequate sensitivity/resolution to survey B-fields in young massive-star forming clouds beyond NGC 6334.

**Acknowledgments**

Our deepest appreciation goes to Ray Blundell for the SMA Director's Discretionary Time that allowed the project to take off. We thanks the help from the SPARO team led by Giles Novak and the Hertz team led by Roger Hildebrand. The discussion with Zhi-Yun Li and the proofread done by David Wilmshurst and Tingting Wu significantly helped the manuscript.

The experiment has been supported by the Hong Kong Research Grants Council, project ECS 24300314, and by the Deutsche Forschungsgemeinschaft priority program 1573, project #46.

The Submillimeter Array is a joint project between the Smithsonian Astrophysical Observatory and the Academia Sinica Institute of Astronomy andAstrophysics and is funded by the Smithsonian Institution and the Academia Sinica.

**Author contributions**

Li designed and executed the experiment. Yuen measured the field curvatures. Otto performed the numerical simulations. The CUHK team is in charge of the manuscript. The CfA-ASIAA-Nanjing team helped with the SMA data acquisition and reduction.

**Contact author information**
H.-b. Li (hbli@phy.cuhk.edu.hk)

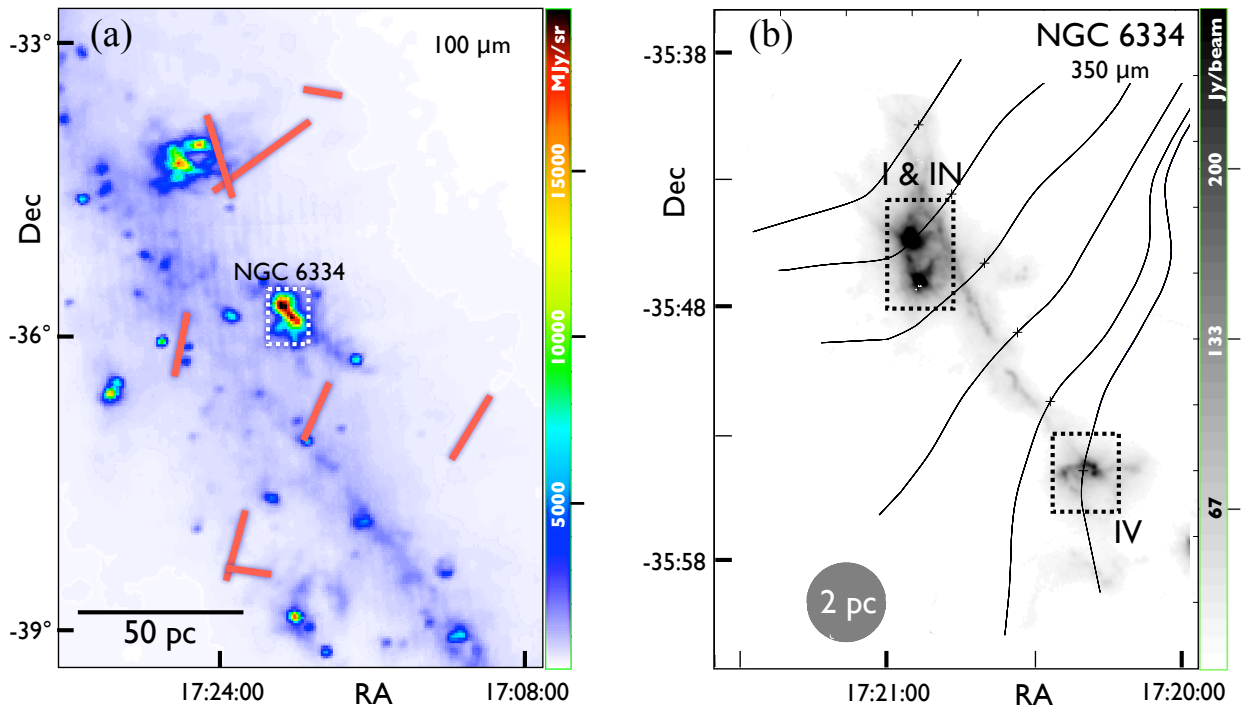

**Figure 1 The B-field directions of NGC 6334 and the local inter cloud medium**
(a): B-field directions (red) inferred from optical polarimetry[15] overlapped with the 100 μm map from IRAS. (b): Zooming in (a), SPARO[13] showed the field lines inferred from the 450 μm polarimetry with a resolution of 2 parsec overlapped with a 350 μm map[30]. *The filamentary cloud "pinches" the field lines and the intensity peaks at the two ends of the filament* (dashed rectangles), where the B-field morphologies with higher resolutions are shown in Fig. 2. The top two field lines are used to estimate the field curvature.

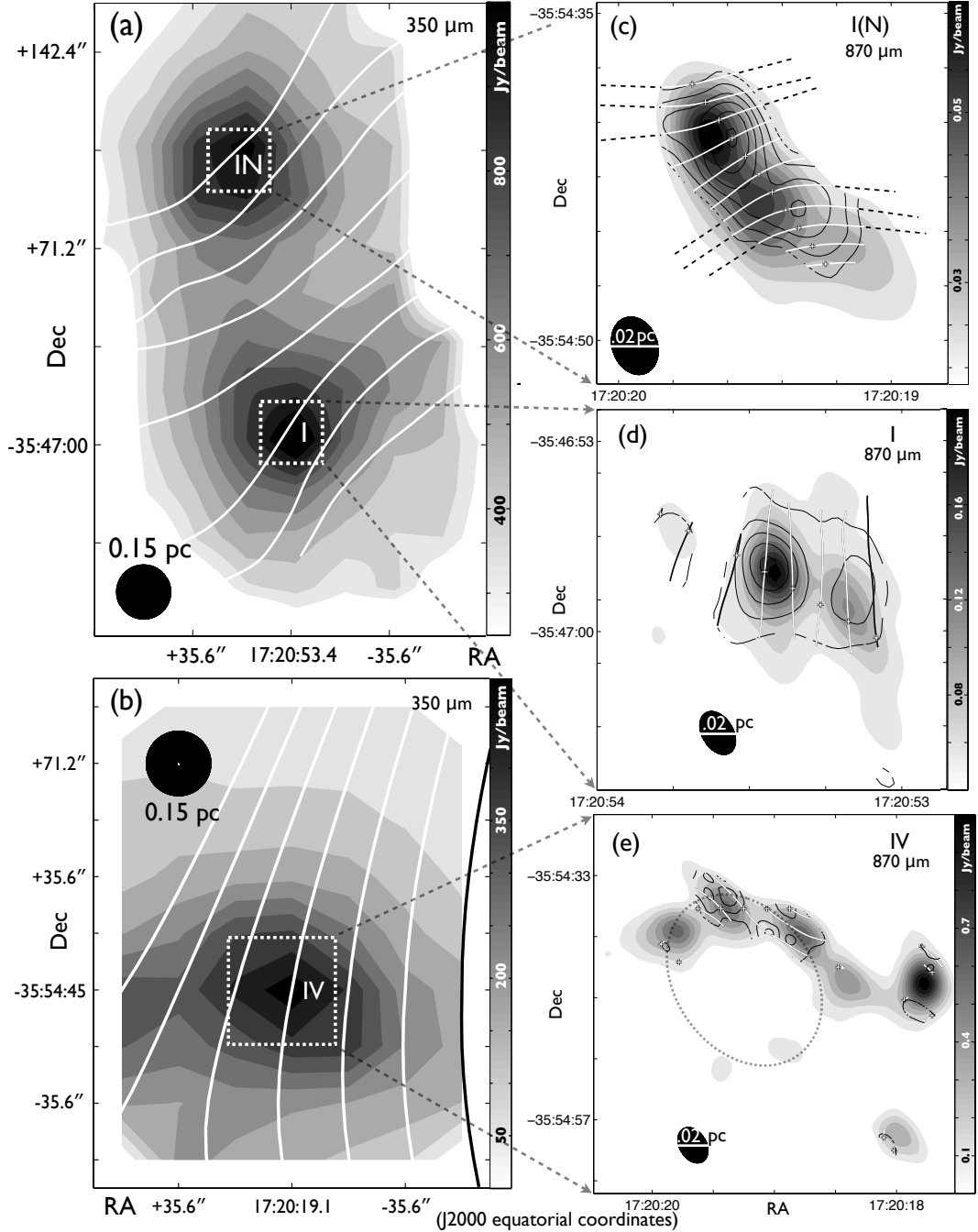

**Figure 2 The B-field within the clumps/cores**

(a) & (b): Clumps observed by Hertz/CSO[18] with a 0.15-parsec resolution. The *intensity again peaks near the field line pinches* in (a). The four field lines passing through the dashed rectangles in (a) are used to estimate the field curvature.

(c)-(e): Cores observed by SMA[6] with a 0.02-parsec resolution. Some field lines in (c) are extended (as dashed lines to help visualizing the pinches) and are used to estimate the curvature. In (e), the oval indicates a shell $H_{II}$ region[20]. The contours show the relative intensity of polarized flux, which tends to increase with the total intensity.

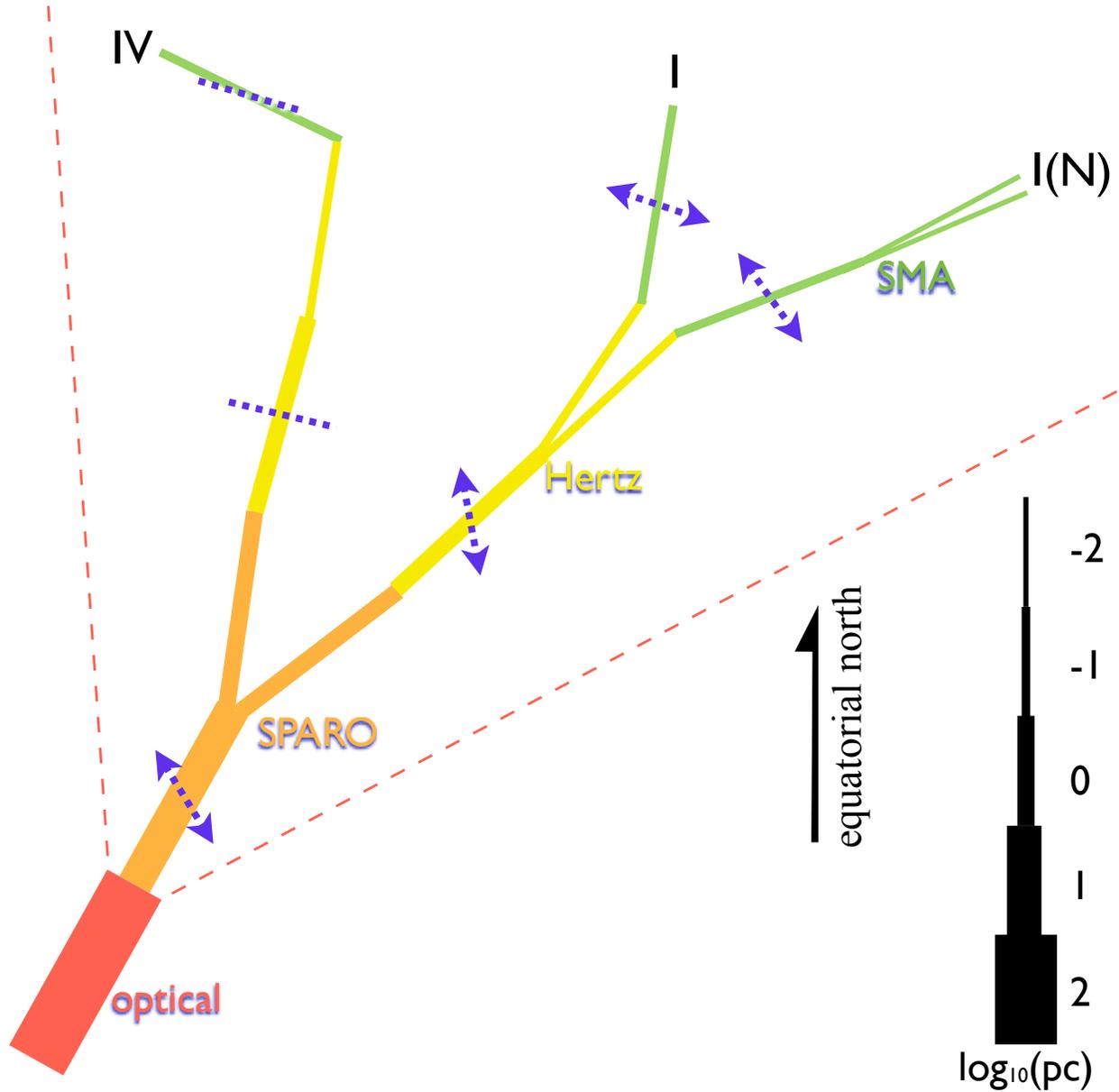

**Figure 3 Self-similar fragmentation and field configurations at 100-0.01 pc.** Each solid line shows the *mean* field direction within a map, whose scale is indicated by the line width (legend in the lower right). The blue dashed lines show the cloud long-axis directions. At the ends of a dashed line, arrowheads are added if the density peaks at the ends of the cloud, where the field directions are indicated by the branched lines. The red dashed lines deviate from the mean inter-cloud-medium field (optical) by 30°. Besides core IV, B-field directions vary within the range defined by the red dashed lines.

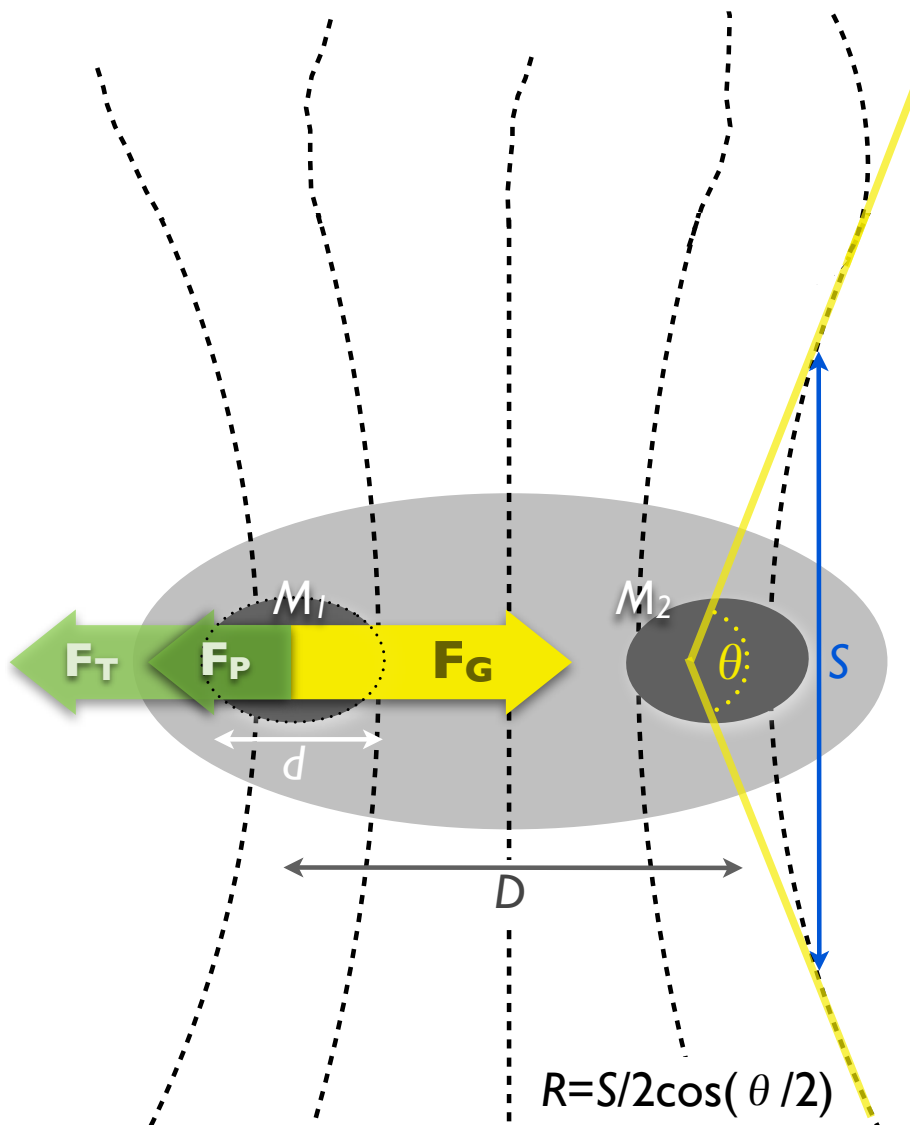

**Figure 4 Parameters used to estimate B-field strength.**
$M_1$ & $M_2$ - clump masses.
$d$ - clump size (for estimating $V$).
$D$ - distance between clumps.
$R$ - field-line (dashed line) curvature radius. From the tip of a pinch, moving in both directions along the field line, one can draw tangents (yellow lines), which form an angle $\theta$. We stop moving the tangent points when $\theta$ stop decreasing. $R$ can then be derived from the equation shown, where $S$ is the separation between the two tangent points.

# Methods

*1. B-field direction alignment between scales*

Note that the *B-field dispersion* test on the relative strength between turbulence and B-field (Fig. 3) might not work when getting even closer to star vicinities, where gas rotation and/or stellar feedback govern the field orientations. The same reason may explain why protostellar disks are found to align with fields at $10^{-2}$ pc scale (poloidal)[33] but not at $10^{-3}$ pc[34]. The transformation from poloidal (cores) to toroidal (discs) fields[35] can orient B-fields in any direction at $10^{-3}$ pc scale.

*2. Estimate of B-field strength*

*2.1 The assumption of $F_G = F_P + F_T$*

We assumed that self-gravity and magnetic fields are close to Virial equilibrium, the so-called critical condition. In practice, we use force equilibrium, $F_G = F_P + F_T$, instead of Virial equilibrium to estimate field strengths; the latter involves volume integration over the cloud. In section 3, we show that our result is roughly consistent with Virial equilibrium. Here we explain why the cloud should be close to critical.

First, given the hourglass-shaped field morphologies, self-gravity is able to compress the fields. So the cloud cannot be significantly subcritical ($F_G \ll F_P + F_T$).

Second, given the elongation of the cores and clumps, the cloud cannot be significantly supercritical ($F_G \gg F_P + F_T$), either. A spherical contraction happens only when the gas is significantly supercritical and it should not be difficult to appreciate that the stronger the field is, the more elongated the clumps/cores should be. To study the relation between cloud elongation and criticality, the ZEUS-MP code[36] is used to simulate cloud contraction. We designed the simulations as simple as possible to focus on the interaction between self-gravity and B-fields. The initial condition is a uniform spherical cloud embedded in a uniform B-field, with negligible gas pressure, no turbulence and no ambipolar diffusion. The only variable is the B-field strength, such that the mass-to-flux ratio (MFR) is 1, 2, 4 and 8 times of the critical value[37]. The short-to-long axis ratios obtained after 10M years of contraction are shown in the Extended Data Fig. 2. The first thing to notice is that it does not take much super-criticality for a nearly spherical contraction; MFR > 4 is enough. Extended Data Fig. 2 also displays the measured axis ratios of the clumps or cores in NGC 6334; indeed they are found to be close to the critical condition

Third, recent studies[3,24] compared the empirical threshold column density of cloud contraction[38,39] with the magnetic critical column density[24,25] of the typical Galactic field (~10 $\mu$G) and found a very good agreement. For densities lower than this threshold, the field strength is independent of density[24, 25], i.e., gas is accumulated along field lines. This is consistent with the scenario of fragmentation channeled by B-fields: sub-critical gas is accumulated along the field lines till the cloud becomes critical and able to compress the field lines (pinches)[3, 24]. The cloud in Fig. 1 (b) indeed looks much more elongated compared to the critical condition in Extended Data Fig. 2. Moreover, due to flux freezing, field-line compression will not change the magnetic criticality[40], and thus magnetic Virial equilibrium should be a good approximation for all scales. In this picture, if Zeeman measurements are used to estimate field strengths, we should expect clouds to *range from just critical to highly supercritical* due to projection effect (because only the LOS components are detected by Zeeman measurements). This range is indeed observed by Crutcher et al.[25]. We would like to emphasize that this interpretation of the Zeeman measurements is debated; other authors[25] interpret the observed range as indication that some cores can be highly supercritical. Highly supercritical, however, is not supported by the surveys mentioned above showing that contraction threshold agrees well with the magnetic critical column density[3,24]. Moreover, it is better realized now that Zeeman measurements can potentially underestimate mean field strengths[24] due to $B_{LOS}$ reversals within a telescope beam[41,42,43,44]. In any case, filaments like NGC 6334 should not belong to the highly-supercritical category even if there is one, because of the low short-to-long axis ratio.

*2.2 Comparison with Chandrasekhar-Fermi Method*

Attributing all the field structures to turbulence, Chandrasekhar & Fermi[3,21,22] proposed to estimate field strength as follows:

$$B = \frac{1}{2}\sqrt{4\pi\rho}\frac{\delta v}{\delta \alpha} \ (Gauss),$$

where $\rho$ is gas density ($g/cm^3$), $\delta v$ is the LOS velocity dispersion ($cm/s$) and $\delta \alpha$ is the B-field direction angle dispersion (*radian*); the factor 1/2 is a correction suggested by numerical simulations[45]. Take Fig. 2 (a), clump I/I(N), as an example, $\delta\alpha$ is measured as 17.5° [46]. The full-width-half-maximum line width of CO(2-1) emission is detected as 13.7 km/s for core I and 12.1 km/s for core I(N) at the 26″ scale (the beam size)[47]. Using the separation between cores I and I(N), 106.8″, as an estimate of the clump size, the line width at the clump scale can be estimated by $(\frac{13.7+12.1}{2})(\frac{106.8}{26})^{0.5} = 26.1$ (km/s), assuming a 0.5 exponent for the turbulent velocity spectrum[48]. Converting the line width to velocity dispersion gives $\delta v$=11.1 (km/s). Assuming $n(H_2)$ = $10^4$ cm$^{-3}$ and a mean molecular mass of 2.8, the above equation gives $B$ = 1.4 m$G$, which is comparable to our estimate of 1.2±0.7 m$G$ (Extended Data Table I). Note that apparently gravity also plays a role in the field structure of Fig. 2 (a) (the hourglass shape), so the estimate from C-F method should be a lower limit.

*3. Mass-to-flux ratio (MFR)*

From Table I, we can roughly estimate the MFR, which is familiar to astronomers when comparing gravitational and magnetic forces. The critical MFR (i.e., when self-gravity and magnetic fields reach Virial equilibrium) is sensitive to cloud geometries; for example, $1/2\pi\sqrt{G}$ for a disc[37] and $2/3\pi\sqrt{G}$ for a spherical cloud[49], where G is the gravitational constant. Assuming *D* from Table I as the cross-section diameters, the MFRs normalized to the critical value are approximately *1.1±0.24*, *2.4±0.74*, *2.2±0.54* and *1.7±0.32* for the cloud, clump I/I(N), core I(N) and core I, respectively, based on the equation from reference 37; correspondingly, the values are *0.83±0.18*, *1.8±0.55*, *1.7±0.41* and *1.3±0.25* based on reference 49. The shapes of our objects are between a disc and a sphere, and the average of the above values is *1.6±0.5*. While the approximations of the cloud shapes and cross-sections are rough, the cloud is unlikely to be highly super-critical. We can check the consistency between the observed MFR and *B-n* relation using the same sets of simulations discussed in section 2.1. The *B-n* relations for the first 10M years are shown in the Extended Data Fig. 2. The 0.4-power indeed occurs when the MFR is 1 to 2 times of the critical value, which is consistent with our observation.

# Extended Data

**Extended Data Table I Parameters used to estimate B-field strength.** Parameters needed to estimate field strength based on $F_G = F_P + F_T$ are listed here. All these parameters are derived assuming 1.7 kiloparsecs to NGC 6334[26], which has a ~25% uncertainty[31,32]. This uncertainty, however, will not affect the estimate of $B$, because $M$ [26] and thus $F_G$ are proportional to the square of distance, so are $F_T$ and $F_P$ (see the equations in main text).

| scale (parsec) | $M_1$ (100M$_\odot$) | $M_2$ (100M$_\odot$) | $D(M_1 M_2)$ (parsec) | clump size $d$ (parsec) | B curvature radius $R$ (parsec) | $B$ [d] (mG) |
|---|---|---|---|---|---|---|
| 10 [a] | 90±18 [26] | 70±14 [26] | 6.0±0.1 [26] | 1.7±0.1 [26] | 1.3±0.45 | 0.19±0.08 |
| 1 | 30±6 [26] | 20±4 [26] | 0.9±0.1 [26] | 0.6±0.1 [26] | 0.49±0.14 | 1.2±0.7 |
| 0.1 (IN) | 0.5±0.25 [27] | 0.5±0.25 [27] | 0.04±0.01 [b] | 0.04±0.01 [c] | 0.05±0.0075 | 13±10 |
| 0.1 (I) | 0.5±0.25 [27] | 0.5±0.25 [27] | 0.05±0.01 [b] | 0.05±0.01 [c] | ∞ | 11±7.5 |

(a) For example, $M_1$ and $M_2$ in the 10-parsec map are defined as follows. The total mass is estimated as $1.6 \times 10^4$ M$_\odot$[26]. $M_1$ is mainly the I/I(N) region (which also defines $d$) with the center of mass at 17:20:53.5 -35:45:22.0 (J2000). The rest of the mass is defined as $M_2$, with the center of mass at 17:20:17.2 -35:54:48.6 (J2000), which is very close to the peak of core IV. The two centers of mass are 6 parsecs apart in the sky.
(b) Defined by the separation between the two intensity peaks in Fig. 2.
(c) Approximated by $D(M_1 M_2)$.
(d) Error estimate: $M$ - 20% and 50% based on references 26 and 27; $D$ and $d$ - based on the beam sizes of references 26 and 27; $R$ - the standard error of the measurements from Figs. 1-2. $B$ - propagated from previous columns.

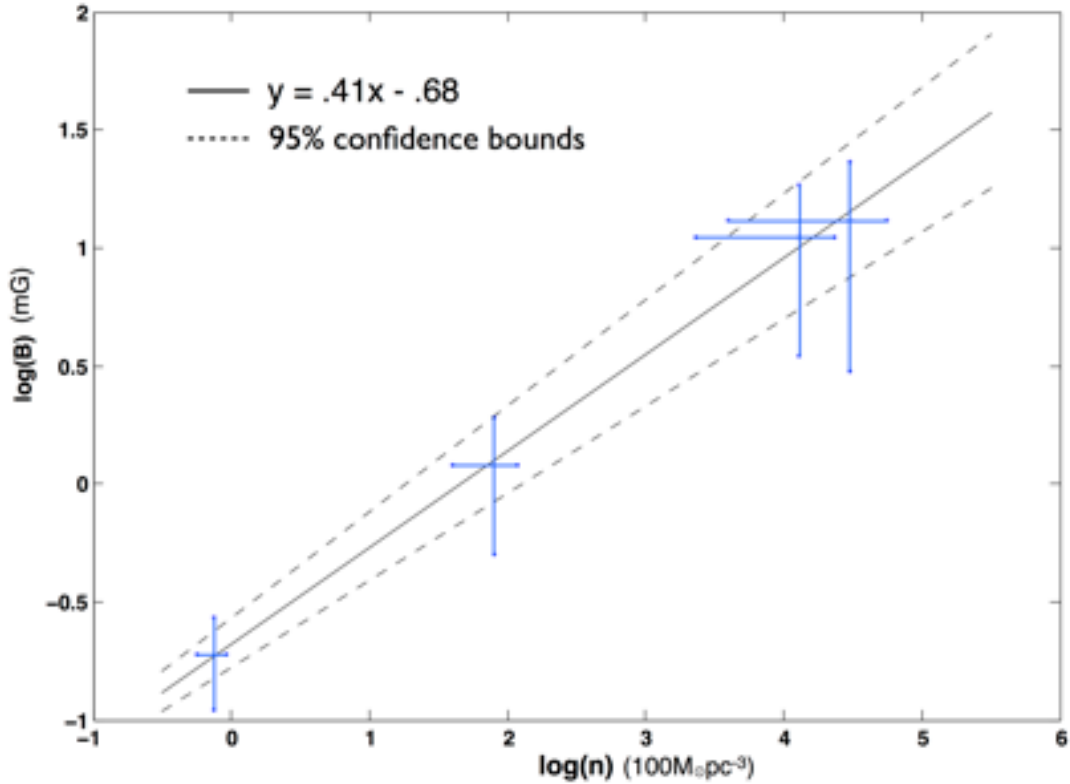

**Extended Data Figure 1 A fitting of the observed *B* and *n* weighted by the signal-to-noise ratio**. *B* is from Table I and *n* is approximated by $(M_1+ M_2)/D^3$. The uncertainties of *M, D, d*, and *R* (Table I) are propagated to *B* and *n*; the error bars of 1-sigma are shown. The slopes of the two dashed lines are .37 and .44. The Curve Fitting toolbox of Matlab is used to fit the data. While projection may affect measurements of field strengths (though not much in our case due to the special LOS), the exponent is less affected, because the effect is the same for all the densities if the field directions are aligned (Fig. 3)

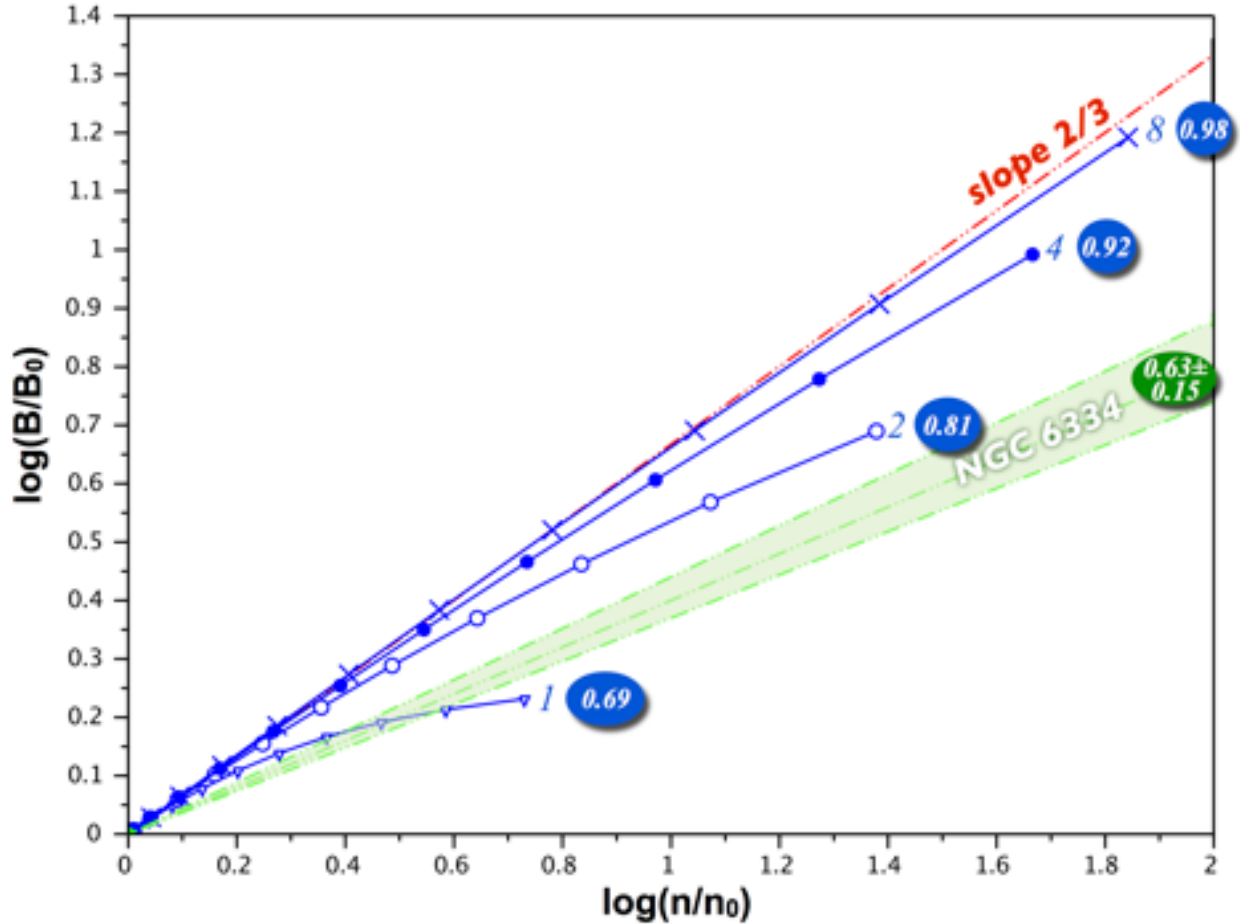

**Extended Data Figure 2 Simulated *B-n* relations and cloud elongation with various magnetic criticality numbers.** Blue lines are the results from the simulations (see text) with various initial uniform field strengths ($B_0$). The initial uniform density of the spherical cloud is $n_0$. $B$ and $n$ are the mean values within a cloud (regions with $n > n_0$). The MFRs normalized by the critical value[45] (criticality numbers) are shown near the ends of each blue line. The slope should never go beyond 2/3 (the red dashed line), the condition of isotropic contraction. A simulation with the criticality number of 600 was also performed and the slope is exactly 2/3. The observed slope of NGC 6334 with a 95% confidence bound is shown by the shaded zone. The short-to-long axis ratio after 10 Myr of each simulation is shown within a blue oval shaped with the same axis ratio. For simplicity, the contour of 20% the peak value is used to define the short and long axes of a cloud. The ratios are measured in the same way for panels (a)-(d) of Fig. 2, and their mean and standard deviation are also shown within the green oval.